\documentclass[twocolumn,floatfix,prl,aps,superscriptaddress]{revtex4}
\usepackage[T1]{fontenc}
\usepackage[latin9]{inputenc}
\usepackage{times}
\usepackage{color} 
\usepackage{xspace}
\usepackage{amssymb,amsmath}
\usepackage{amsbsy}
   \usepackage{dsfont}
\usepackage{graphicx}
\usepackage{bm}
\usepackage{epstopdf}
\usepackage{float}
\usepackage{adjustbox}

\usepackage{soul}

\usepackage[unicode,breaklinks]{hyperref}
\hypersetup{
    unicode=true,
    a4paper=true,
    plainpages=false, 
    colorlinks=true,
    linkcolor=blue,
    citecolor=blue,
    filecolor=black,
    urlcolor=blue
}
\begin{document}
 \title{Catalyzation of supersolidity in binary dipolar condensates}
\author{D. Scheiermann}
\affiliation{Institut f\"ur Theoretische Physik, Leibniz Universit\"at Hannover, Germany}
\author{L. A. Pe\~na Ardila}  \thanks{luis.ardila@itp.uni-hannover.de}
\affiliation{Institut f\"ur Theoretische Physik, Leibniz Universit\"at Hannover, Germany}
\author{T. Bland}
\affiliation{Institut f\"ur Quantenoptik and Quanteninformation, Innsbruck, Austria}
\affiliation{Institut f\"ur Experimentalphysik, Universit\"at Innsbruck, Austria}
\author{R. N. Bisset}
\affiliation{Institut f\"ur Experimentalphysik, Universit\"at Innsbruck, Austria}
\author{L. Santos} \thanks{santos@itp.uni-hannover.de}
\affiliation{Institut f\"ur Theoretische Physik, Leibniz Universit\"at Hannover, Germany}

\begin{abstract}
Breakthrough experiments have newly explored the fascinating physics of dipolar quantum droplets and supersolids.
The recent realization of dipolar mixtures opens further intriguing possibilities.  
We show that under rather general conditions, the presence of a second component catalyzes droplet nucleation and supersolidity in an 
otherwise unmodulated condensate. Droplet catalyzation in miscible mixtures, which may occur even for a surprisingly small impurity doping,  
results from a local roton instability triggered by the doping-dependent modification of the effective dipolar strength. The catalyzation mechanism 
may trigger the formation of a two-fluid supersolid, characterized by a generally different superfluid fraction of each component, which 
opens intriguing possibilities for the future study of spin physics in dipolar supersolids.
\end{abstract} 
  
\maketitle



Supersolids constitute an intriguing state of matter that combines superfluidity and the crystalline 
order characteristic of a solid~\cite{Boninsegni2012}. Whereas this long-sought phase has remained elusive in Helium~\cite{Chan2013}, 
recent developments on ultra-cold gases have opened new possibilities for its realization. 
Bose-Einstein condensates with spin-orbit coupling~\cite{Li2017,Putra2020, Geier2021} and in optical cavities~\cite{Leonard2017} have revealed supersolid features. 
Recently, breakthrough experiments on condensates formed by highly magnetic atoms have created dipolar supersolids with an interaction-induced 
crystalline structure~\cite{Review2021,Review2022}.

Dipolar supersolids are closely linked to the idea of quantum droplets, a novel ultra-dilute quantum liquid that results from  
the combination of competing, and to large extend cancelling, mean-field interactions, and the stabilization 
provided by quantum fluctuations~(quantum stabilization)~\cite{Petrov2015}. In dipolar condensates~\cite{Review2021,Review2022}, the competition between
strong dipolar interactions and contact-like interactions provides the crucial mean-field quasi-cancellation.  
Quantum stabilization arrests mean-field collapse leading to self-bound 
droplets~\cite{Kadau2016,Chomaz2016,Schmitt2016}. Due to the anisotropy and non-locality of the dipolar interaction, 
confinement leads  to the formation of a droplet array~\cite{Wenzel2017}, which for properly fine-tuned contact interactions remains 
superfluid, hence realizing a supersolid~\cite{Tanzi2019, Boettcher2019, Chomaz2019, Natale2019, Tanzi2019b, Guo2019}. 
The recent creation of two-dimensional dipolar supersolids~\cite{Norcia2021, Bland2021} opens further fascinating perspectives, as exotic 
pattern formation~\cite{Zhang2019, Zhang2021, Hertkorn2021, Poli2021} and quantum vortices~\cite{Gallemi2020,Roccuzzo2020}.

Quantum droplets have been also realized in non-dipolar binary mixtures due to competing mean-field 
intra- and inter-component contact interactions and quantum stabilization~\cite{Cabrera2018, Semeghini2018, DErrico2019}.  Crucially, droplet formation requires miscible components 
with a fixed ratio between their densities given by the ratio of intra-component scattering lengths~\cite{Petrov2015}. 
 Hence, the mixture behaves as a single-component condensate. Moreover,  the short-range isotropic character 
 of the interactions prevents the formation of droplet arrays and supersolids, although self-bound supersolid stripes may be realized in the presence of 
 spin-orbit coupling~\cite{Sachdeva2020,Sanchez-Baena2020}.



\begin{figure} [t!]
\begin{center}
\includegraphics[width=0.7\columnwidth]{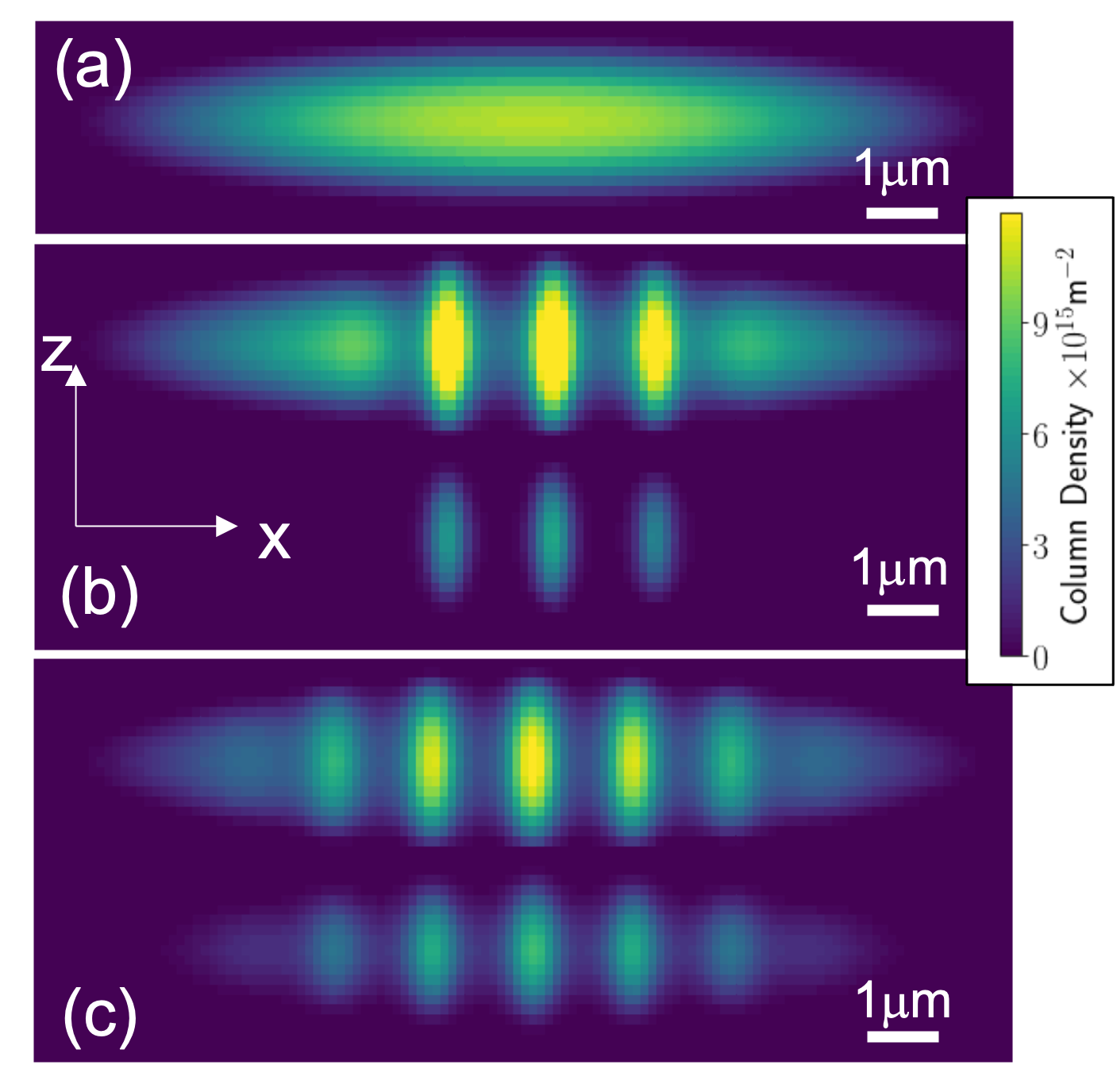}
\caption{Binary mixture with $N=63000$ $^{164}$Dy atoms, 
$\omega_{x,y,z}=2\pi(33,110,167)\, \mathrm{Hz}$, $\mu_1=10\mu_B$, $\mu_2=9\mu_B$, 
and $a_{11}=a_{22}=100a_0$. 
(a) Column density (integrated along $y$) for $N_2=0$. (b) Column density of component 1 (top) and 2 (bottom) for $a_{12}=62.5a_0$ and $N_2/N=0.1$~(SS-ID regime).
(c) Same for $a_{12}=67.5a_0$ and $N_2/N=0.3$~(SS-SS regime). In Figs.~(b) and (c) the components are shifted along $z$ for visualization purposes.}
 \label{fig:1}
\end{center}
\end{figure}


Recent experiments have realized for the first time mixtures of two dipolar species~\cite{Trautmann2018, Durastante2020, Politi2021}, opening 
exciting new possibilities. In contrast to non-dipolar mixtures, dipolar droplets 
may occur for an arbitrary relative population of the components, which may be both miscible and immiscible~\cite{Bisset2021, Smith2021, Smith2021b}. 
Interestingly, the presence of a second dipolar component may lead to the formation of droplets. 
Recent numerical results have shown that the increase of inter-species contact
interactions in an Er-Dy mixture in the presence of gravitational sag may lead to droplet nucleation in one of the components~\cite{Politi2021, footnote-Modugno}. 

In this Letter, we discuss, how doping an unmodulated condensate with a second miscible component induces a local modification of the relative dipolar strength, 
which may result, even for a tiny doping, in a local roton instability that triggers droplet nucleation and supersolidity in both components.  The resulting two-fluid supersolid, characterized 
by a generally different superfluid character of each component, constitutes a rich novel scenario that opens intriguing possibilities for the future study of  
the interplay between the density and the spin degrees of freedom in dipolar supersolids.



\paragraph{Model.--}  
We consider two bosonic components, $\sigma = \{1,2\}$, formed by magnetic atoms~(a similar formalism applies to electric dipoles). 
The components may be either two different atomic species, as e.g. erbium and dysprosium~\cite{Trautmann2018, Durastante2020, Politi2021}, 
or atoms of the same species in different internal states, e.g. $^{164}$Dy in different spin states, which is the case we employ below to illustrate the possible physics. 
Using the formalism of Ref.~\cite{Bisset2021}, we evaluate the Lee-Huang-Yang~(LHY) energy density in an homogeneous binary mixture 
with densities $n_{1,2}$ (assuming equal masses $m_{1,2}=m$):
\begin{equation}
\xi_{\mathrm {LHY}} (n_{1,2})\!=\!\! \frac{16}{15\sqrt{2\pi}}\! \left ( \frac{m}{4\pi\hbar^2}\right )^{\!\!\frac{3}{2}}\! {\cal R}\! \left [\!\int_0^1 \!\!\!du\!\sum_{\lambda=\pm} V_\lambda(u,n_1,n_2)^{\!\frac{5}{2}} \!\right ], 
\label{eq:LHY_2comp}
\end{equation}
with ${\cal R}[ \, ]$ the real part, and 
\begin{equation}
V_\pm (u,n_1,n_2) \!=\!\! \sum_{\sigma=1,2}\!\!\eta_{\sigma\sigma}n_\sigma \pm \sqrt{(\eta_{11}n_1\!-\!\eta_{22}n_2)^2\!+\!4\eta_{12}^2n_1 n_2},
\end{equation}
where  $\eta_{\sigma\sigma'}(u)=g_{\sigma\sigma'}+g_{\sigma\sigma'}^d (3u^2-1)$. The contact-like 
interactions between components $\sigma$ and $\sigma'$ are characterized by the coupling constant 
$g_{\sigma\sigma'}\equiv \frac{4\pi\hbar^2 a_{\sigma\sigma'}}{m}$, with $a_{\sigma\sigma'}$ the corresponding $s$-wave scattering length. The strength of the 
dipole-dipole interactions between components $\sigma$ and $\sigma'$ is given by 
$g_{\sigma\sigma'}^d \equiv \frac{\mu_0\mu_\sigma\mu_{\sigma'}}{3} \equiv \frac{4\pi\hbar^2 a_{\sigma\sigma'}^{d}}{m}$, where $\mu_\sigma$ is the magnetic dipole moment 
of component $\sigma$, and $\mu_0$ is the vacuum permeability. 
Below we fix $\mu_1=10\mu_B$ and $\mu_2=9\mu_B$, with $\mu_B$ the Bohr magneton. 
In order to study spatially inhomogeneous dipolar mixtures we apply the local-density approximation to the LHY term, obtaining the energy functional:
\begin{eqnarray}
E &=& \int d^3 r \left [ \sum_\sigma \psi_\sigma^*(\vec r)\left ( \frac{-\hbar^2\nabla^2}{2m} + V_{\mathrm {trap}}(\vec r)  \right ) \psi_\sigma(\vec r)  \right\delimiter 0 \nonumber \\
&+&  \frac{1}{2} \sum_{\sigma,\sigma'} g_{\sigma\sigma'} n_{\sigma}(\vec r)n_{\sigma'}(\vec r)  + \xi_{\mathrm {LHY}}[n_{1}(\vec r),n_{2}(\vec r)] \nonumber \\
&+&\left\delimiter 0 \sum_{\sigma,\sigma'} \frac{3g_{\sigma\sigma'}^{d}}{8\pi} 
\int d^3 r' U_{dd}(\vec r-\vec r') n_{\sigma'}(\vec r\,')  n_\sigma(\vec r) \right ] , 
\label{eq:E_2comp}
\end{eqnarray}
with $V_{\mathrm {trap}}(\vec r)=\frac{m}{2}\left ( \omega_x^2 x^2+\omega_y^2 y^2 + \omega_z^2 z^2\right )$, 
$n_\sigma(\vec r)=|\psi_\sigma(\vec r)|^2$, and $U_{dd}(\vec r)=\frac{1}{r^3}(1-3\cos^2\theta)$, 
with $\theta$ the angle between $\vec r$ and the dipole moments. The ground-state of a mixture with $N_{\sigma}$ atoms in component $\sigma$, with 
a total atom number $N=N_1+N_2$,  
is obtained from the coupled extended Gross-Pitaevskii equations~(eGPE): 
\begin{equation}
\tilde\mu_\sigma \psi_\sigma(\vec r)=\frac{\partial E}{\partial \psi_\sigma^*(\vec r)},
\end{equation} 
where $\int d^3 r \, n_\sigma(\vec r)=N_\sigma$ and $\tilde\mu_\sigma$ is the chemical potential of component $\sigma$. 
We consider below an elongated trap, with $\omega_{x,y,z}= 2\pi\times(33,110,167)\,\mathrm{Hz}$,
and $N=63000$ atoms. Similar values have been employed in recent experiments~\cite{Norcia2021}.



\begin{figure} [t!]
\begin{center}
\includegraphics[width=\columnwidth]{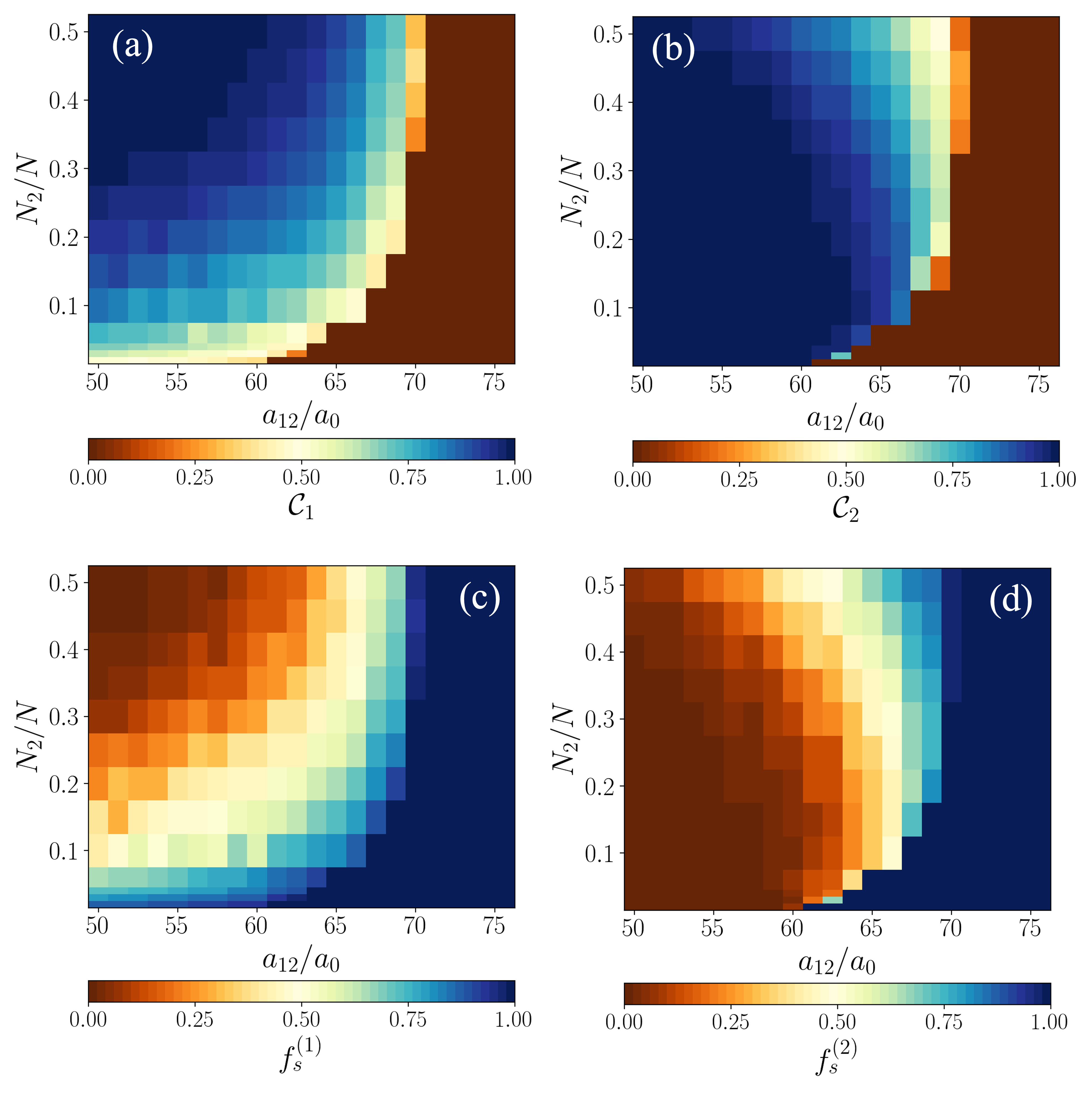}
\caption{(a,b) Contrast of components 1 and 2, respectively, as a function of $a_{12}/a_{11}$ and $N_2/N$. 
(c,d) Superfluid fraction of components 1 and 2, respectively. The parameters are the same as in Fig.~\ref{fig:1}.}
 \label{fig:2}
\end{center}
\end{figure}




\paragraph{Doping-induced supersolidity.--} 
Single-component dipolar condensates in elongated traps present three possible ground-states: the unmodulated~(U) regime, without density
modulation, at large-enough scattering length $a$; the incoherent-droplet (ID) regime, an incoherent linear droplet at low-enough
$a$; and the supersolid~(SS) regime, a coherent droplet array in a narrow intermediate window of values of $a$.
 
In order to illustrate the new possibilities for droplet nucleation and supersolidity in dipolar mixtures, 
we consider in the following $a_{11}=100a_0$ such that, in absence of component 2, component $1$ is well in the U regime~(Fig.~\ref{fig:1}(a)), 
which demands $a_{11}>94a_0$.  Although below we provide a more general picture, we fix at this point $a_{22}=a_{11}$ for simplicity. 
We evaluate the ground-state of the mixture as a function of the doping $N_2/N$ and $a_{12}$.
We focus on the case in which  $a_{12}$ is low-enough, such that the mixture remains miscible. 

For a sufficiently large $a_{12}$~($>72a_0$ in the case considered) the binary mixture remains unmodulated (U regime). 
In contrast, for  $a_{12}< 72a_0$ and a sufficiently large $N_2/N$, droplet nucleation leads to a density modulation in both components.  
The spatial modulation of the density profile of component $\sigma$ is best characterized using the 
contrast, determined from the maximal and minimal density in the central region ($|x|<L=2\mu \mathrm{m}$ in our calculations), $n_{\sigma,\mathrm {max}}$ and $n_{\sigma,\mathrm {min}}$, as ${\cal C}_\sigma=(n_{\sigma,\mathrm {max}}-n_{\sigma,\mathrm {min}})/(n_{\sigma,\mathrm {max}}+n_{\sigma,\mathrm {min}})$. The contrast for components 1 and 2 is depicted in Figs.~\ref{fig:2}~(a) and~(b). 
The superfluid fraction in each component, $f_{s}^{(\sigma)}$, may be estimated using Legget's upper bound in the central window~\cite{Legget1970,Legget1998}:
 
\begin{equation}
f_{s}^{(\sigma)} = (2 L)^2 \left [ \int_{-L}^{L} dx \tilde n_\sigma(x) \int_{-L}^{L} dx \frac{1}{\tilde n_\sigma(x)} \right ]^{-1},
\end{equation}
with $\tilde n_\sigma(x)=\iint\, dy\, dz\, n_\sigma(\vec r)$. For components of equal mass,
the overall superfluid fraction, $f_s=\frac{N_1}{N}f_s^{(1)}+\frac{N_2}{N}f_s^{(2)}$, 
characterizes the reduction of the overall momentum of inertia,  
as would be measured via e.g. the response to a component-independent scissors-like perturbation~\cite{Tanzi2021, Norcia2021b, Roccuzzo2022, footnote-CounterMotion}.

For sufficiently low $a_{12}$, both components form incoherent droplet arrays~(ID-ID regime). The contrasts approach their maximum
${\cal C}_{1,2}\simeq 1$, and $f_s^{(1,2)}\simeq 0$.  As for other scenarios of dipolar droplets~\cite{Tanzi2019, Boettcher2019, Chomaz2019}, 
the contrasts are significantly lower~(and $f_s^{(1,2)}$ significantly higher) close to the transition into the U regime, marking the SS-SS regime, in which the droplets 
of both components remain coherently linked. Note, that ${\cal C}_{1}$ and ${\cal C}_{2}$ follow a different dependence on $N_2/N$. 
Droplets in component 1~(2) may remain coherent, while those in component 2~(1) are in the ID regime, determining the SS-ID (ID-SS) regimes~(see Fig.~\ref{fig:1}(b) and~\cite{footnote-SM}). 
Defining the ID regime as having a contrast ${\cal C}>0.96$~\cite{footnote-C}, we determine the ground-state diagram of Fig.~\ref{fig:3}~(for a discussion of the overall superfluidity and 
the relative contribution to each component to it, see~\cite{footnote-SM}). Interestingly, for low-enough $a_{12}$, a surprisingly small $N_2/N$ is enough to induce droplet nucleation.



\begin{figure} [t!]
\begin{center}
\includegraphics[width=0.8\columnwidth]{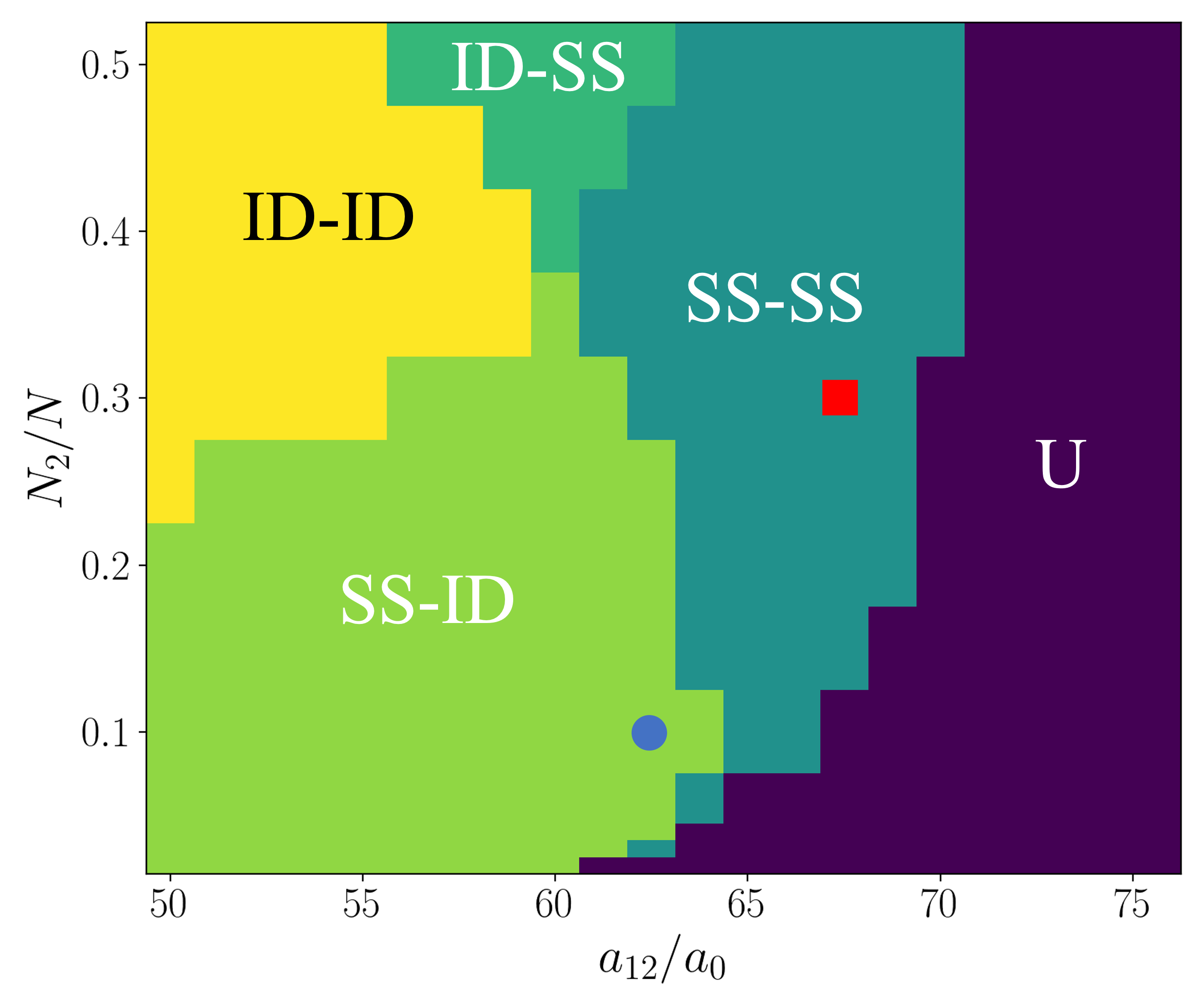}
\caption{Ground-state phases of the binary mixture as a function of $a_{12}/a_0$ and $N_2/N$, for the same parameters as Fig.~\ref{fig:1}. 
The blue circle and red square indicate the cases of Figs.~\ref{fig:1}~(b) and~(c), respectively.}
 \label{fig:3}
\end{center}
\end{figure}



\begin{figure} [t!]
\begin{center}
\includegraphics[width=\columnwidth]{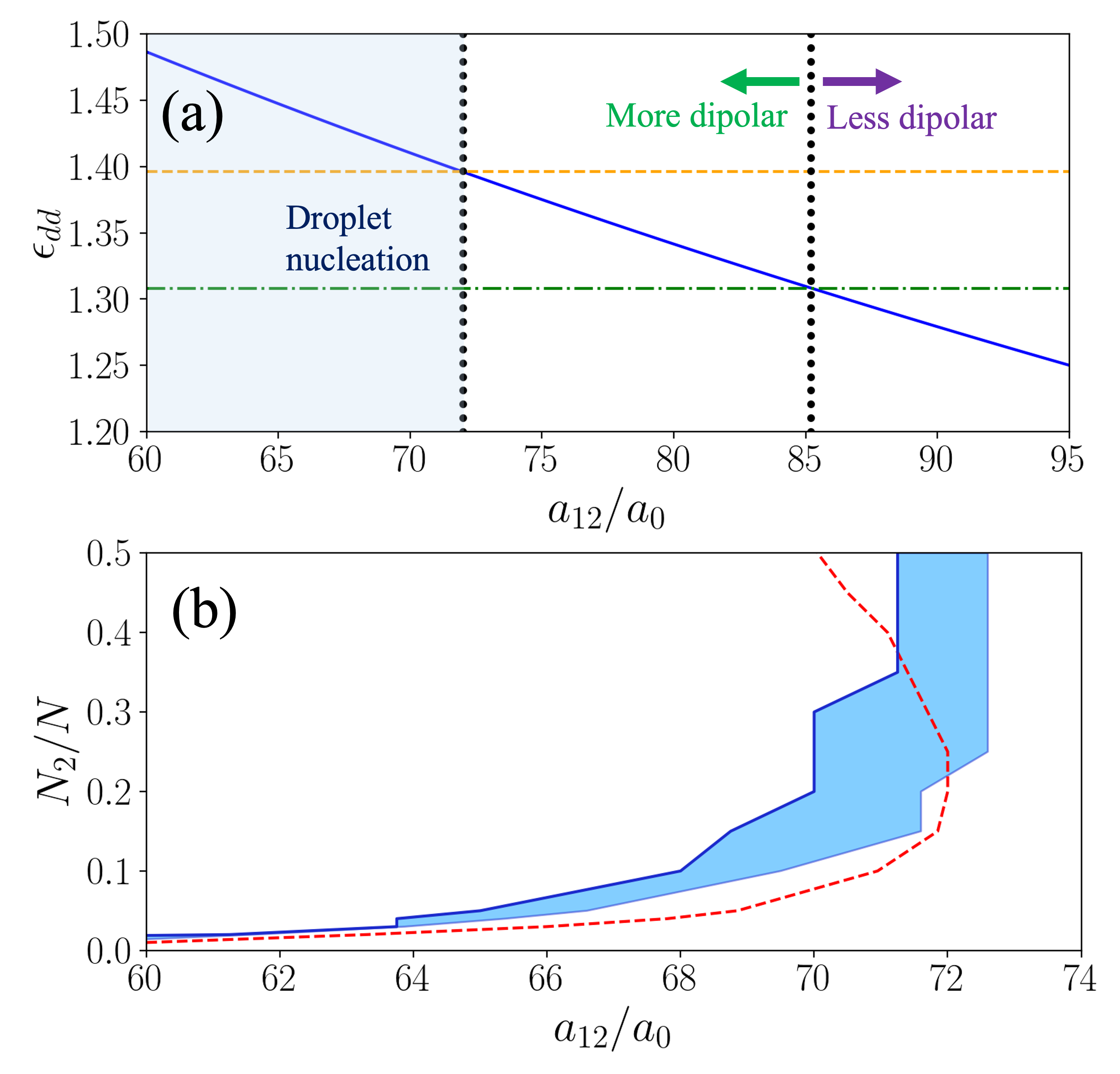}
\caption{(a) Effective dipolar strength $\epsilon_{\mathrm{dd}}$ (blue-solid curve) for the parameters of  Fig.~\ref{fig:1} and $P=1$, as a function of $a_{12}$. 
For $a_{12}<85a_0$, $\epsilon_{\mathrm{dd}}(P=1)>\epsilon_{\mathrm{dd}}(P=0)$~(dot-dashed line), and hence the mixture is more dipolar than the component 1 alone. 
For  $a_{12}<72a_0$, $\epsilon_{\mathrm{dd}}>1.39$~(dashed line), the value at which droplets start to be nucleated in a single component. (b) Onset of the density modulation. 
The solid dark-blue curve depicts for a given $N_2/N$ the value of $a_{12;cr}$ , such that for $a_{12}<a_{12;cr}$ both components have a non-zero contrast. 
The blue-shaded area indicates the threshold regime of $a_{12}$ values for which we observe at the trap center the growth of a significant deviation from the unmodulated density profile, although no local density minima have yet developed, and hence the contrast $\cal{C}$ remains zero. The red/solid curve indicates the onset of the central roton-instability as obtained from Eq.~\eqref{eq:DeltaR}, using the central density $n_0$ and polarization $P$ evaluated in the U regime for $a_{12}=75\, a_0$.}
 \label{fig:4}
\end{center}
\end{figure}

\paragraph{Effective dipolar strength and local roton instability.-- } We introduce at this point a model, that shows that droplet catalyzation is triggered  
by a roton instability due to the local modification of the effective dipolar strength. 
The unmodulated mixture presents an approximately constant polarization $n_2(\vec r)/n_1(\vec r) \simeq n_2(0)/n_1(0)=P$ in the central 
region in which the density modulation develops at the U-SS transition.  This is satisfied well for a balanced mixture, but it remains a fairly good approximation 
down to the impurity regime~\cite{footnote-SM}. We may then introduce the local-polarization approximation, reducing our analysis to the case of 
a mixture with constant polarization $P$, where
$\psi_1(\vec r)\simeq \frac{1}{\sqrt{1+P}} \psi(\vec r)$ and $\psi_2(\vec r)\simeq \sqrt{\frac{P}{1+P}} \psi(\vec r)$. This reduces the problem to a scalar model 
given by the energy functional: 
 \begin{eqnarray}
E &=& \int d^3 r \left [ \psi(\vec r)^* \left ( \frac{-\hbar^2\nabla^2}{2m} + V_{\mathrm {trap}}(\vec r) \right ) \psi(\vec r) \right\delimiter 0 \nonumber \\
&+& \frac{1}{2} g_{\mathrm{eff}}(P) \chi(P)  n(\vec r)^2     \nonumber \\
&+& \left\delimiter 0 \frac{3g_{\mathrm{eff}}^d(P)}{8\pi} 
\int d^3 r' U_{dd}(\vec r-\vec r')  n(\vec r\, ')n(\vec r)    \right ]. 
\label{eq:E_1comp}
\end{eqnarray}
where $n(\vec r)=|\psi(\vec r)|^2$, $g_{\mathrm{eff}}(P)=4\pi\hbar^2 a_{\mathrm{eff}}(P)/m$, and $g_{\mathrm{eff}}^d(P) = 4\pi\hbar^2 a_{\mathrm{eff}}^d(P)/m$. Note that 
for the relatively low densities that characterize the U regime, we have included (to a very good approximation~\cite{footnote-SM}) the LHY contribution to the chemical potential
 by regularizing $g_{\mathrm{eff}}\to g_{\mathrm{eff}}\chi$, with 
 \begin{equation}
\chi\! =\! 1 + \frac{8\sqrt{n_0a_{\mathrm{eff}}^3}}{3\sqrt{2\pi}}  {\cal R}\! \left [ \int_0^1 \!\!du\! 
\!\sum_{\lambda=\pm}\! \left [\frac{V_\lambda \left ( u, \frac{1}{1+P}, \frac{P}{1+P} \right )}{g_{\mathrm{eff}}}\right ]^{\frac{5}{2}} \right ].
\end{equation}
Crucially, the system is characterized by a polarization-dependent effective scattering length $a_{\mathrm{eff}}(P) = \frac{a_{11}+P^2a_{22}+2 P a_{12}}{(1+P)^2}$, 
and dipolar length $a_{\mathrm{eff}}^d(P) = \frac{a_{11}^d+P^2a_{22}^d+2 P a_{12}^d}{(1+P)^2}$.  
As a result the effective dipolar strength $\epsilon_{\mathrm{dd}}=a_{\mathrm{eff}}^d/a_{\mathrm{eff}}$ depends on the local polarization. 
In Fig.~\ref{fig:4}(a) we depict $\epsilon_{\mathrm{dd}}$ for $P=1$ (which is approximately the central polarization for $N_2/N=0.5$ in our calculations), as a function of $a_{12}$. For $a_{12}<85\, a_0$~(for the parameters of Fig.~\ref{fig:1}) the mixture is locally effectively more dipolar than the component 1 alone, i.e. 
$\epsilon_{\mathrm{dd}}(P=1)>\epsilon_{\mathrm{dd}}(P=0)$. For a sufficiently low $a_{12}$, $\epsilon_{\mathrm{dd}}$ is large enough to drive the system locally unstable, as discussed below.



Since the condensate is very elongated in the axial direction, we may approximate the density profile of the unmodulated mixture at the axial trap center as $n(\vec r)\simeq n(y,z)$. 
Assuming a transversal Thomas-Fermi profile,  $n(y,z) \simeq n_0 \left ( 1 - y^2/R_y^2 - z^2/R_z^2 \right )$, 
we may directly employ the formalism developed in Ref.~\cite{Chomaz2018} for the study of the roton instability in an 
elongated single-component dipolar condensate~\cite{footnote-SM}. 
The roton energy $\Delta_{\mathrm{R}}$ acquires the form:
 \begin{eqnarray}
\left (\frac{ \Delta_{\mathrm{R}}}{g_{\mathrm{eff}} n_0} \right )^2 &=&  \sum_{j=y,z} \Lambda_j^2 f_j(\kappa,\epsilon_{\mathrm{dd}})
- \frac{4}{9}\left (\epsilon_{\mathrm{dd}}-\chi \right )^2
\label{eq:DeltaR}
 \end{eqnarray}
where $\Lambda_j = \frac{\hbar\omega_j}{g_{\mathrm{eff}}n_0}$, 
$f_j(\kappa,\epsilon_{\mathrm{dd}})=\frac{\epsilon_{\mathrm{dd}}(1+\kappa)^2/2}{(1+\kappa)^2\chi+\epsilon_{\mathrm{dd}}(2+\beta_j\kappa-\kappa^2)}$, $\beta_y=-2$, $\beta_z=4$, 
and the transversal aspect ratio $\kappa=R_z/R_y$ is given by $(\omega_y/\omega_z)^2 = \kappa^2 f_z(\kappa,\epsilon_{\mathrm{dd}})/f_y(\kappa,\epsilon_{\mathrm{dd}})$. 

From the eGPE calculations, we obtain $n_0=n(0)$ and $P$ for different $N_2/N$ and $a_{12}=75\, a_0$, in the U regime close to the 
SS transition. From the condition $\Delta_{\mathrm{R}}=0$, we determine the 
value of $a_{12}$ that , according to the model above, would result in roton instability at the center of the unmodulated condensate.
The results are in very good qualitative agreement (and indeed also quantitative, considering the deviations from both the local-polarization approximation and the transversal Thomas-Fermi profile) with the threshold for droplet nucleation obtained from the coupled eGPEs~(Fig.~\ref{fig:4}(b)). 
Droplet catalyzation hence results from the roton instability induced by the locally-modified dipolar strength.



\paragraph{Conclusions.--} Doping with a second dipolar component may catalyze droplet nucleation and supersolidity in an otherwise unmodulated dipolar condensate. 
This stems from the local modification of the effective dipolar strength, which for a sufficiently low value of $a_{12}$ results in a local roton instability, which 
triggers droplet nucleation, once arrested by quantum stabilization. We note that droplet catalyzation does not 
require full overlapping of both components, but only an overlapping region with an extension larger than the transversal size of the droplet.
Strikingly, catalyzation may be efficient even for a very small doping, well within the impurity regime.

Droplet catalyzation allows for the realization of a two-fluid supersolid, characterized 
by generally different superfluid fractions in each component.  We stress that, although the triggering mechanism behind droplet nucleation may be understood from a locally-valid scalar model~\eqref{eq:E_1comp}, 
the supersolid mixture that forms once droplets nucleate is a genuinely two-component system, and not an effective 
scalar condensate. 
The two-fluid supersolid thus constitutes a novel platform for the study of the interplay between density modulation, 
spin physics, and the different superfluid character of each component.  The intriguing properties of two-fluid dipolar supersolids, and in particular their 
elementary excitations, will be the focus of future research. 



\begin{acknowledgments}
We thank V. Cikojevic for his contribution during the first stages of this work. 
We acknowledge support of the Deutsche Forschungsgemeinschaft (DFG, German Research Foundation) under Germany's Excellence Strategy -- EXC-2123 QuantumFrontiers -- 390837967, 
and FOR 2247.  T. B. acknowledges funding from FWF Grant No. I4426  2019.
\end{acknowledgments}





\newpage

\renewcommand{\theequation}{S\arabic{equation}}
\renewcommand{\thefigure}{S\arabic{figure}}
\renewcommand{\thetable}{S\arabic{table}}


\setcounter{equation}{0}
\setcounter{figure}{0}
\setcounter{table}{0}

\onecolumngrid
\noindent\rule{18cm}{0.4pt}

\section*{SUPPLEMENTAL MATERIAL:\\ "Catalyzation of supersolidity in binary dipolar condensates"}
\begin{center}
D. Scheiermann,$^{1}$ 
Luis A.~Pe{\~n}a~Ardila,$^{1}$  
T. Bland,$^{2,3}$ 
R. N. Bisset$^{3}$ and
L. Santos $^{1}$ \\
\emph{\small $^1$Institut f\"ur Theoretische Physik, Leibniz Universit\"at Hannover, Germany}\\
\emph{\small $^2$Institut f\"ur Quantenoptik and Quanteninformation, Innsbruck, Austria}\\
\emph{\small $^3$ Institut f\"ur Experimentalphysik, Universit\"at Innsbruck, Austria}\\
\end{center}

In this Supplemental Material, we provide additional examples of the different ground states phases discussed in the main text, and make additional comments on
 the overall superfluidity and the relative contribution of each component to it. We discuss as well 
further details concerning the simplified model employed in the main text for the understanding of the droplet catalyzation. 
Finally, we briefly comment on the possibility of realizing a supersolid of immiscible droplets.\\


\section{Change of the density profiles of the components}

\begin{figure} [H]
\begin{center}
\vspace{0.4cm}
\includegraphics[width=\columnwidth]{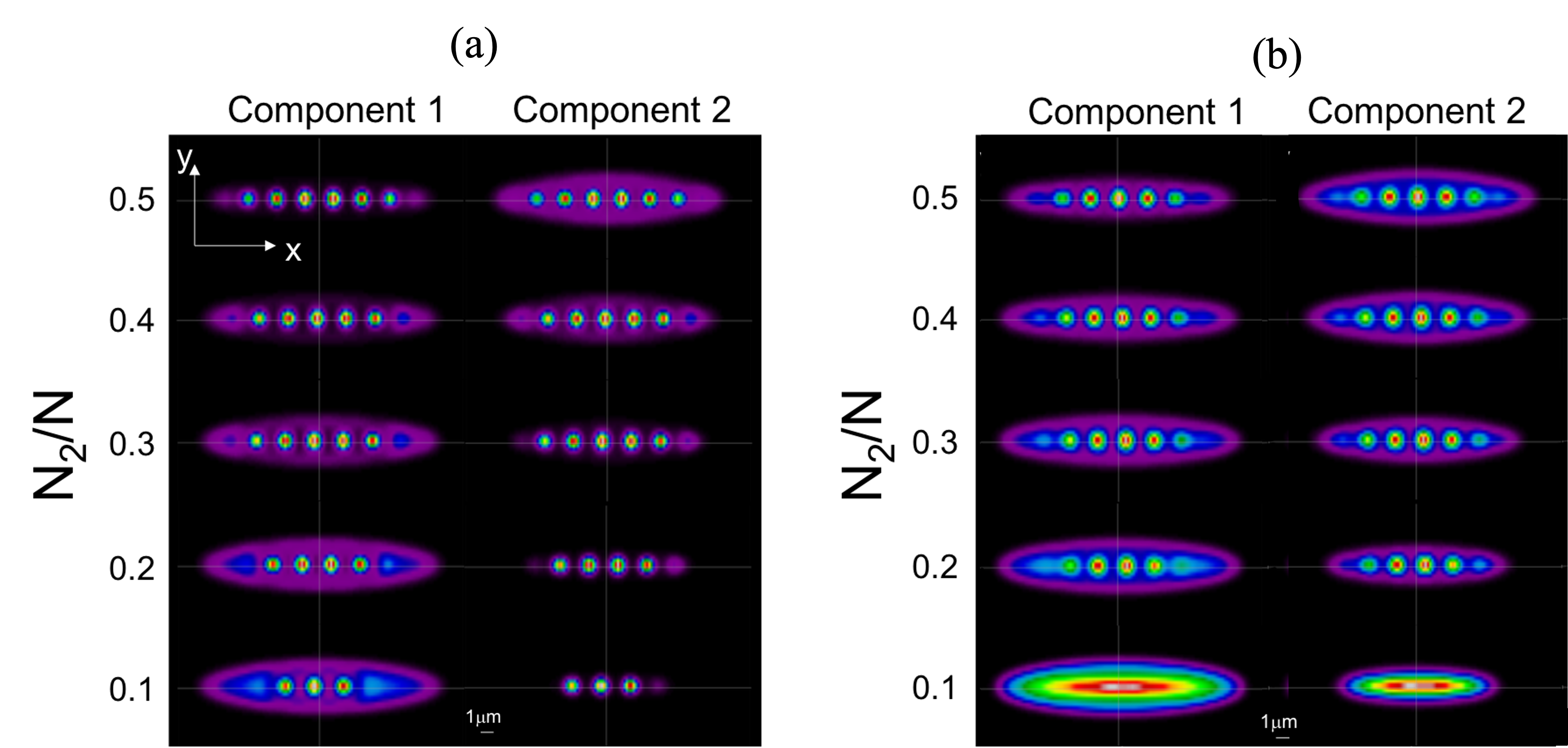}
\caption{(a) Binary mixture with $N=63000$, 
$\omega_{x,y,z}=2\pi(33,110,167)\,\mathrm{Hz}$, $\mu_1=10\mu_B$, $\mu_2=9\mu_B$, 
$a_{11}=a_{22}=100a_0$ and $a_{12}=62.5a_0$. (b) Same but with $a_{12}=67.5a_0$}
 \label{fig:S1}
\end{center}
\end{figure}

Figure 3 of the main text shows the possible ground-states of the miscible mixture: U, SS-SS, SS-ID, ID-SS, and ID-ID.  
In Figs. 1(b) and (c) of the main text we present two cases, in the SS-ID and SS-SS regimes, respectively.
In Figs.~\ref{fig:S1}~(a) and~(b), we illustrate in more detail the transition experienced by the density profiles as a function of $N_2/N$.
Figure~\ref{fig:S1}~(a) shows the density profiles of both components for the same parameters as those of Fig. 1 of the main text, for the case of $a_{12}=62.5a_0$. 
For growing $N_2/N$ the system transitions from the SS-ID regime to the SS-SS regime and finally into the ID-SS regime.  
In Fig.~\ref{fig:S1}~(b) we depict the density profiles of both components for the same case but with $a_{12}=67.5a_0$. For growing $N_2/N$ the system transitions 
from the U regime into the SS-SS regime.

Concerning our numerical calculations, we have solved the coupled eGPE equations using standard split-operator and fast-Fourier transformation techniques. We employ in the 
calculations shown in the paper numerical boxes $|x|<12 l_x$, $|y|<6l_x$, $|z|<6l_x$, with $l_x=\sqrt{\hbar/m\omega_x}=1.36\,\mu\mathrm{m}$, and 
a number of points $N_{x,y,z}=128, 64, 32$. We have checked that increasing the boxes and/or the number of points does not change appreciably our results.
 


\section{Overall superfluidity}

\begin{figure} [H]
\begin{center}
\vspace{0.4cm}
\includegraphics[width=0.7\columnwidth]{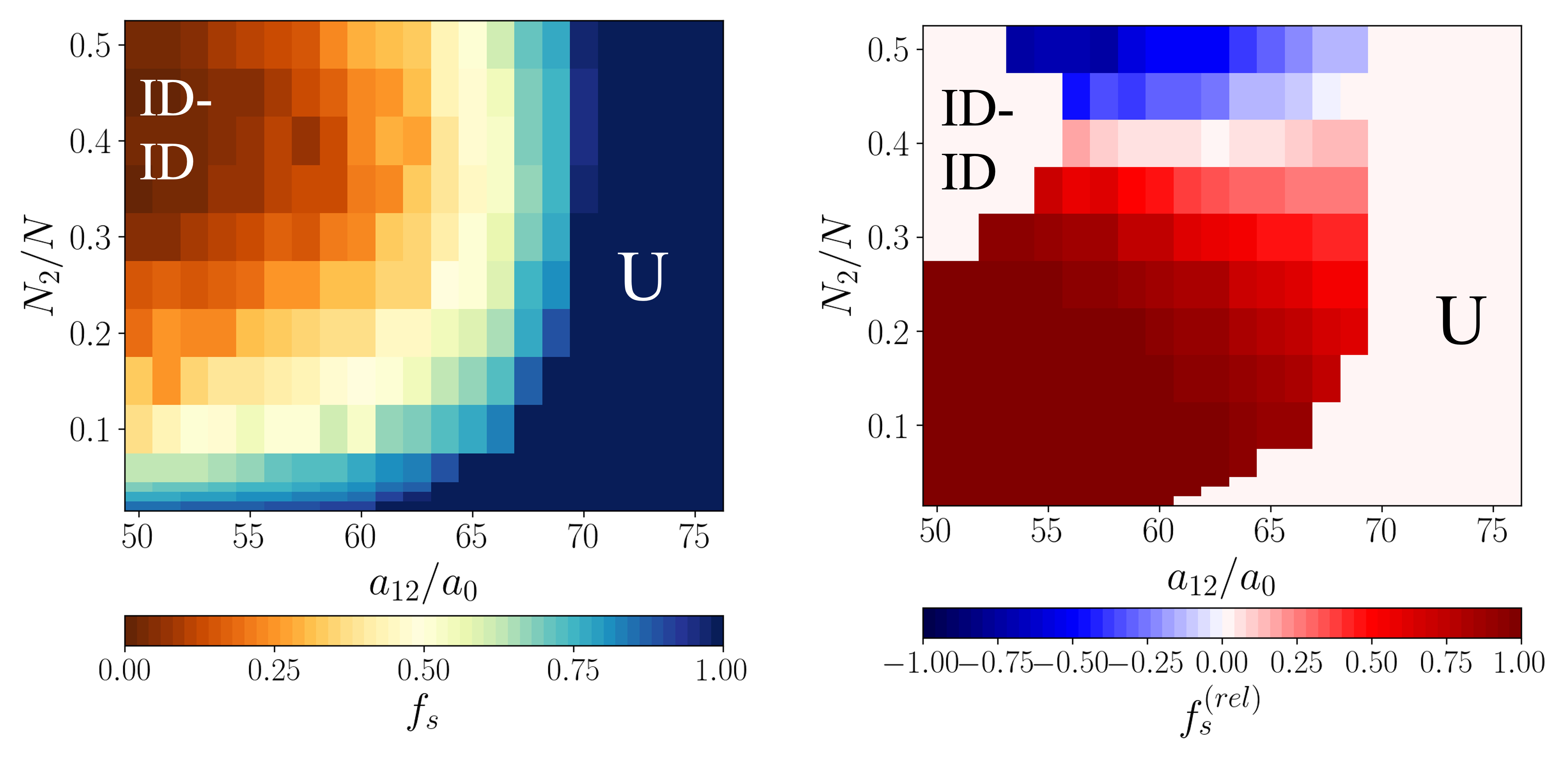}
\caption{Overall superfluid fraction $f_s$~(a) and relative fraction $f_s^{(rel)}$~(b) for the same parameters as Fig.~2 of the main text. In Fig.~(b) we have excluded the U and ID-ID regions. Note that for growing $N_2/N$ the superfluidity transitions from being dominated by the first component to be dominated by the second one. In the SS-ID~(ID-SS) phase 
the superfluidty is to a good approximation only given by component 1~(2).}
 \label{fig:S2}
\end{center}
\end{figure}

As mentioned in the main text, the overall superfluid fraction $f_s=\frac{N_1}{N}f_s^{(1)}+\frac{N_2}{N}f_s^{(2)}$~(see Fig.~\ref{fig:S2}~(a)), characterizes the reduction of the 
momentum of inertia (compared to its classical value) as would be measured e.g. from the response of the mixture to a spin-independent scissors-like perturbation. 
The relative contribution of each component to the overall superfluid fraction may be evaluated from the relative superfluid fraction~(see Fig.~\ref{fig:S2}~(b)):
$f_s^{(rel)}=\frac{1}{f_s}\left ( \frac{N_1}{N}f_s^{(1)}-\frac{N_2}{N}f_s^{(2)} \right )$. In accordance to Fig. 3 of the main text, we see that in addition to the U and ID-ID regimes, there is an intermediate regime with finite overall contrast and superfluidity. Note that whereas for low $N_2/N$ the superfluidity is dominated by the first component, the 
situation reverses when $N_2/N$ approaches $0.5$. In the region we denote as SS-ID~(ID-SS) in Fig. 3 of the main text, superfluidity is given only by component 1 (2).



\section{Evaluation of the local roton instability}

In this section we provide additional details concerning the model employed in the main text for the study of the local roton instability 
that induces the observed droplet catalyzation.


\subsection{Validity of the local polarization approximation}

\begin{figure} [H]
\begin{center}
\vspace{0.4cm}
\includegraphics[width=\columnwidth]{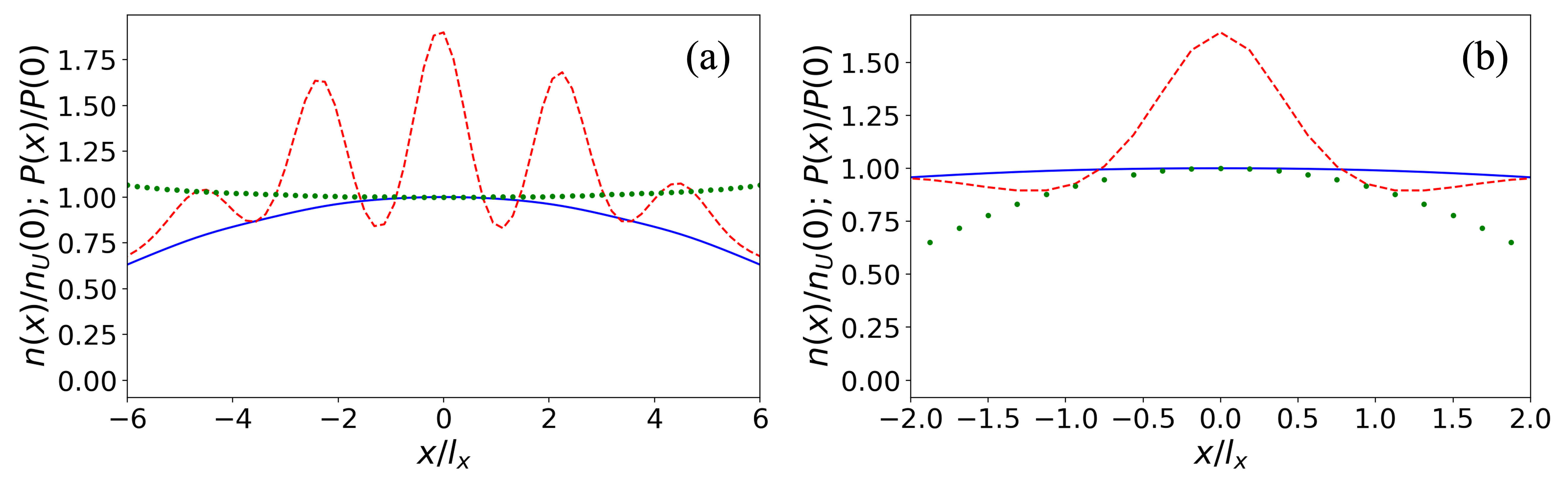}
\caption{(a) Central region where droplets develop. We consider the same parameters as in Fig. 1 of the main text. 
Figure (a) depicts the case of a balanced mixture~($N_2/N=0.5$). The solid-blue and dashed-red curve indicate the density $n_1(x,0,0)$ for $a_{12}=75a_0$~(unmodulated regime) 
and $a_{12}=70a_0$~(modulated regime), respectively, whereas the dotted-green curve depicts the polarization $P=n_2(x,0,0)/n_1(x,0,0)$.
Figure (b) shows the case of a very imbalanced mixture~($N_2/N=0.05$). The solid-blue and dashed-red curve indicate the density $n_1(x,0,0)$ for $a_{12}=66.25a_0$~(unmodulated regime) and $a_{12}=57.5a_0$~(modulated regime), respectively, whereas the dotted-green curve depicts the polarization $P=n_2(x,0,0)/n_1(x,0,0)$.
In both figures, the length unit is $l_x=\sqrt{\hbar/m\omega_x}$, and we have normalized the density by the central value $n_U(0)$ of the density of the unmodulated case, and 
the polarization by its central value $P(0)$.}
 \label{fig:S3}
\end{center}
\end{figure}

The model discussed in the main text employs the fact that at the trap center, in the region where the density modulation develops, 
 the polarization is approximately constant. This is satisfied well in the balanced mixture case, $N_1=N_2$~(see Fig.~\ref{fig:S3}(a)), were the two components have very 
 similar density profiles, and indeed the mixture can be well understood using single-mode approximation in the whole cloud~\cite{Smith2021b}. 
 In contrast, single-mode approximation obviously fails 
 in the impurity regime, since the minority component moves into the central region of the majority component. However, even in that case, the 
 polarization remains approximately constant in the central region where a droplet develops~(see Fig.~\ref{fig:S3}(b)).



\subsection{Regularization of the contact term}

Assuming a constant polarization $P$, we re-express $n_1=\frac{n}{1+P}$ and $n_2=\frac{nP}{1+P}$, with $P=n_2/n_1$ and $n=n_1+n_2$.
We may then rewrite the contribution of the contact interactions and the LHY to 
the energy functional of Eq. (3) of the main text in the form:
\begin{equation}
E_{\mathrm{SR}+\mathrm{LHY}}(\vec r) = \frac{1}{2} g_{\mathrm{eff}}(P) n(\vec r)^2 + \xi_{\mathrm{LHY}} \left ( \frac{1}{1+P}, \frac{P}{1+P} \right ) n(\vec r)^{5/2}.
\end{equation}
The contribution to the effective extended Gross-Pitaevskii equation for $\psi(\vec r)$ will be given by the local chemical potential:
\begin{equation}
\mu_{\mathrm{SR}+\mathrm{LHY}}(\vec r) =  g_{\mathrm{eff}}(P) n(\vec r) + \frac{5}{2}\xi_{\mathrm{LHY}} \left ( \frac{1}{1+P}, \frac{P}{1+P} \right ) n(\vec r)^{3/2}.
\end{equation}

\begin{figure} [H]
\begin{center}
\vspace{0.4cm}
\includegraphics[width=0.5\columnwidth]{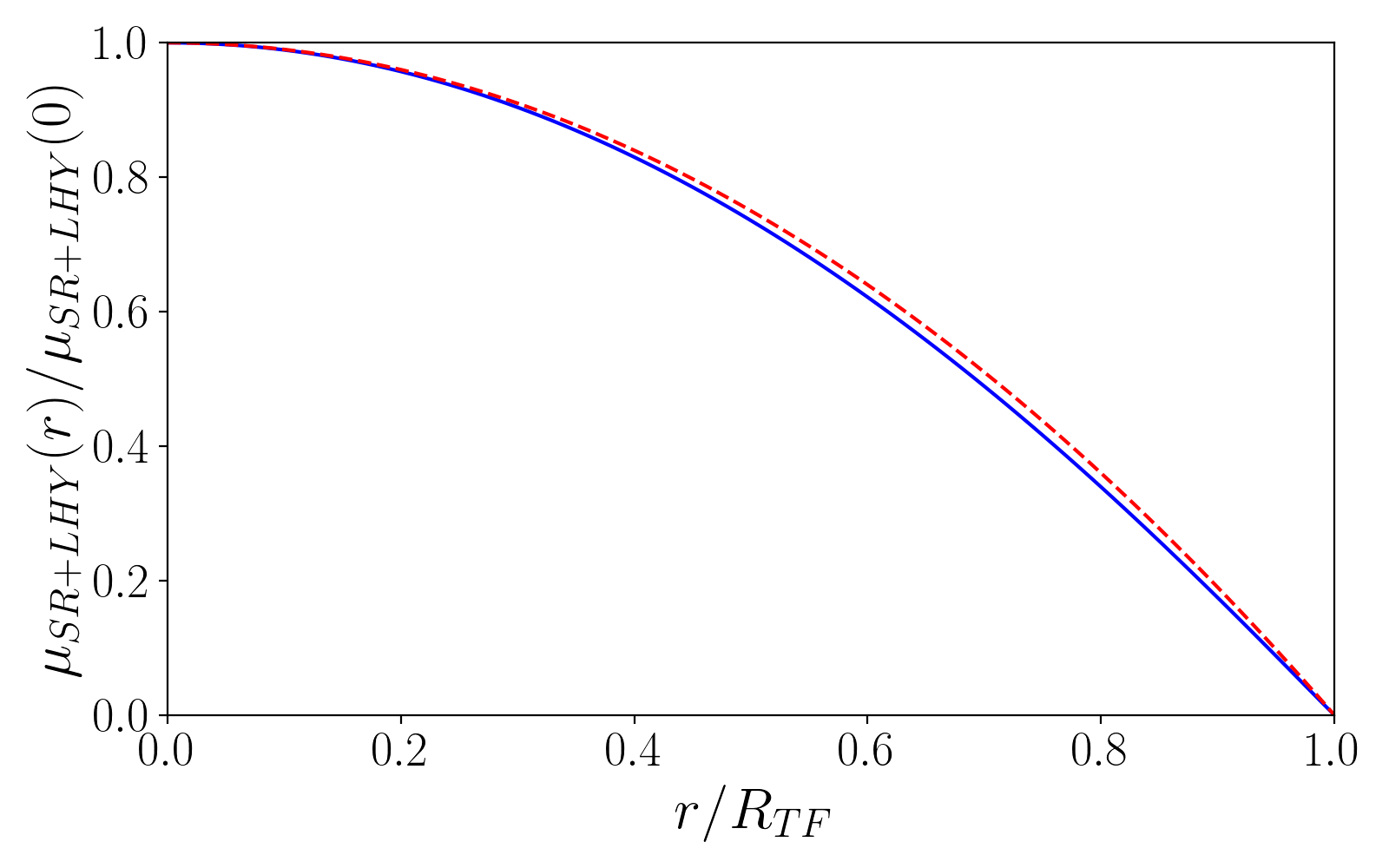}
\caption{Comparison of $\mu_{SR+LHY} [n(r)]$ (solid, blue) and the approximate $g_{\mathrm{eff}}\chi(P) n(r)$~(dashed, red) for a one-dimensional Thomas-Fermi profile $n(r)=n_0 \left (1-r^2/R_{\mathrm{TF}}^2 \right )$. We have employed a density $n_0=5\times 10^{20}\mathrm{m}^{-3}$, and Dy atoms with $a_{11}=a_{22}=100a_0$, $a_{12}=70a_0$, $\mu_1=10\mu_B$, and 
$\mu_2=9\mu_B$, and polarization $P=1$.}
 \label{fig:S4}
\end{center}
\end{figure}

As shown in Fig.~\ref{fig:S4}, this expression may be well approximated by:
\begin{equation}
\mu_{\mathrm{SR}+\mathrm{LHY}}(\vec r) \simeq  g_{\mathrm{eff}}(P) \chi(P) n(\vec r)
\end{equation}
with 
\begin{equation}
\chi(P) = 1 + \frac{5}{2} n_0^{1/2} \frac{\xi_{\mathrm{LHY}} \left ( \frac{1}{1+P}, \frac{P}{1+P} \right ) } {g_{\mathrm{eff}}(P)}  
=  1 + \frac{8}{3\sqrt{2\pi}} \sqrt{n_0 a_{\mathrm{eff}}^3(P)} \,\, {\cal R}\! \left [ \int_0^1 du \sum_{\lambda=\pm} \left ( \frac{V_\lambda\left ( u, \frac{1}{1+P},\frac{P}{1+P}\right)}{g_{\mathrm{eff}}(P)}\right )^{5/2} \right ]
\end{equation}



\subsection{Transversal Thomas-Fermi profile}
Assuming an axially-homogeneous condensate, with a transversal Thomas-Fermi profile, $n=n_0(1-y^2/R_y^2-z^2/R_z^2)$, we may proceed as in Refs.~\cite{Eberlein2005, Chomaz2018} to determine the transversal  Thomas-Fermi radii:
\begin{eqnarray}
R_y^2 &=& \frac{2g_{\mathrm{eff}} (P)?n_0}{m\omega_y^2} \left [ \chi(P) + \epsilon_{\mathrm{dd}}(P) \left ( \frac{2-2\kappa-\kappa^2}{(1+\kappa)^2} \right ) \right ], \label{eq:Ry2} \\
R_z^2 &=&  \frac{2g_{\mathrm{eff}} (P)?n_0}{m\omega_z^2} \left [ \chi(P) + \epsilon_{\mathrm{dd}}(P) \left ( \frac{2+4\kappa-\kappa^2}{(1+\kappa)^2}\right ) \right ], \label{eq:Rz2} 
\end{eqnarray}
with $\kappa=R_z/R_y$, given by the relation:
\begin{equation}
\left ( \frac{\omega_y}{\omega_z} \right )^2 = \kappa^2 \left [ \frac{\chi(P) + \epsilon_{\mathrm{dd}}(P) \left ( \frac{2-2\kappa-\kappa^2}{(1+\kappa)^2} \right )}{\chi(P) + \epsilon_{\mathrm{dd}}(P) \left ( \frac{2+4\kappa-\kappa^2}{(1+\kappa)^2} \right )} \right ]
\end{equation}


\subsection{Roton energy}
Using the procedure outlined in the Suppl. Material of Ref.~\cite{Chomaz2018}, we may determine the excitation spectrum in the vicinity of the roton minimum~(see Eq.~(1) 
in Ref.~\cite{Chomaz2018}):
\begin{equation}
\xi(k_x)^2 \simeq \Delta_{\mathrm{R}}^2 + \frac{2\hbar^2 k_{\mathrm{R}}^2}{m} \frac{\hbar^2}{2m} \left ( k_x -k_{\mathrm{R}} \right )^2,
\end{equation}
where $k_{\mathrm R}$ is the roton momentum and 
\begin{equation}
\Delta_{\mathrm{R}}^2 =2g_{\mathrm{eff}}n_0 \epsilon_{\mathrm{dd}} \frac{\hbar^2}{2m}\left ( \frac{1}{R_y^2} + \frac{1}{R_z^2 }\right ) -\frac{4}{9} \left (g_{\mathrm{eff}}(P) n_0 \right )^2 \left ( \epsilon_{\mathrm{dd}}-\chi(P) \right)^2. 
\end{equation}
Using Eqs.~\eqref{eq:Ry2} and~\eqref{eq:Rz2}, we recover the expression of Eq. (8) of the main text. 



\section{Supersolids of immiscible droplets.}
 
In the main text, we have only focused on the miscible regime. When $a_{12}>0$ increases, inter-component repulsion eventually leads to immiscibility, i.e. to the spatial separation of the components. 
In a non-dipolar gas, immiscibility would result, for the elongated geometry considered, in phase separation along the trap axis for a sufficiently large $a_{12}$. 
This is observed as well when extending the regimes of Figs. 2 and 3 of the main text into larger $a_{12}$ values. A dipolar mixture may open, however, more intriguing scenarios.
We briefly illustrate in the following one of these scenarios.
 
Recent studies~\cite{Smith2021, Bisset2021} have shown that dipolar mixtures allow for immiscible self-bound droplets, in which one of the components 
is attached, along the dipole axis, to the ends of a droplet of the second component. This attachment results from the formation at those locations of energy minima  
induced by the inter-component dipole-dipole interaction. Interestingly, immiscible dipolar mixtures may form a supersolid array of these immiscible droplets. 
This is in particular the case if $a_{22}$ is low-enough, such that the second component alone would nucleate droplets. This is illustrated in Fig.~\ref{fig:3}, where we depict 
the ground-state densities for $a_{11}=95a_0$, $a_{22}=76a_{0}$, and $a_{12}=84a_{0}$, and $N_2/N=0.5$. Note the formation of immiscible droplets, similar as those discussed 
in Refs.~\cite{Bisset2021,Smith2021,Smith2021b}, which arrange in a peculiar double supersolid.



\begin{figure}[tb!]
\begin{center}
\includegraphics[width=0.4\columnwidth]{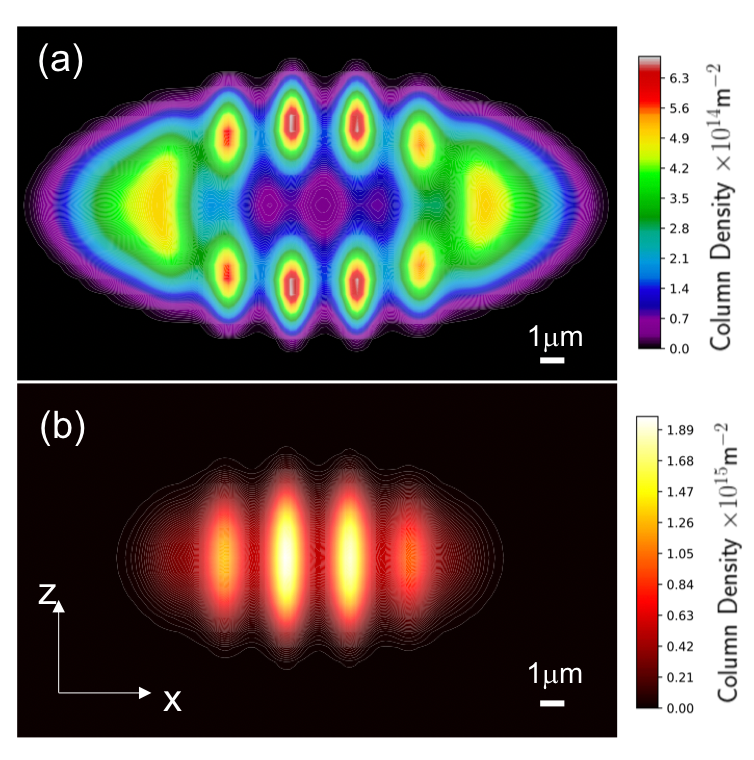}
\caption{Supersolid of immiscible droplets. Column density on the $xz$ plane for the component 1~(a) and 2~(b) for $N=63000$, $(\omega_x,\omega_y,\omega_z)=(2\pi(33,110,167)$Hz, 
$N_2/N=0.5$, $a_{11}=95a_0$, $a_{22}=76a_{0}$, and $a_{12}=84a_{0}$.}
 \label{fig:3}
\end{center}
\end{figure}



\begin{thebibliography}{0}
\expandafter\ifx\csname natexlab\endcsname\relax\def\natexlab#1{#1}\fi
\expandafter\ifx\csname bibnamefont\endcsname\relax
  \def\bibnamefont#1{#1}\fi
\expandafter\ifx\csname bibfnamefont\endcsname\relax
  \def\bibfnamefont#1{#1}\fi
\expandafter\ifx\csname citenamefont\endcsname\relax
  \def\citenamefont#1{#1}\fi
\expandafter\ifx\csname url\endcsname\relax
  \def\url#1{\texttt{#1}}\fi
\expandafter\ifx\csname urlprefix\endcsname\relax\def\urlprefix{URL }\fi
\providecommand{\bibinfo}[2]{#2}
\providecommand{\eprint}[2][]{\url{#2}}

\end{thebibliography}


\begin{thebibliography}{99}

\bibitem{Boninsegni2012} M. Boninsegni and N. V. Prokof'ev, Rev. Mod. Phys. {\bf 84} 759 (2012).

\bibitem{Chan2013} M. H. W. Chan, R. B. Hallock, and L. Reatto, J. Low Temp. Phys. {\bf 172}, 317 (2013).

\bibitem{Li2017} J.-R. Li, J. Lee, W. Huang, S. Burchesky, B. Shteynas, F. C. Top, A. O. Jamison, and W. Ketterle, Nature {\bf 543}, 91 (2017).

\bibitem{Putra2020} A. Putra, F. Salces-C\'arcoba, Y. Yue, S. Sugawa, and I. B. Spielman, Phys. Rev. Lett. {\bf 124}, 053605 (2020).

\bibitem{Geier2021} K. T. Geier, G. I. Martone, P. Hauke, and S. Stringari, Phys. Rev. Lett. {\bf 127}, 115301 (2021).

\bibitem{Leonard2017} J. L\'eonard, A. Morales, P. Zupancic, T. Esslinger, and T. Donner, Nature {\bf 543}, 87 (2017).

\bibitem{Review2021} F. B\"ottcher, J.-N. Schmidt, J. Hertkorn, K. S. H. Ng, S. D. Graham, M. Guo, T. Langen, and T. Pfau, Rep. Prog. Phys. {\bf 84} 012403 (2021).

\bibitem{Review2022} L. Chomaz, I. Ferrier-Barbut, F. Ferlaino, B. Laburthe-Tolra, B. L. Lev,  and T. Pfau,  arXiv:2201.02672.

\bibitem{Petrov2015} D. S. Petrov, Phys. Rev. Lett. {\bf 115}, 155302 (2015).

\bibitem{Kadau2016} H. Kadau, M. Schmitt, M. Wenzel, C. Wink, T. Maier, I. Ferrier-Barbut, and T. Pfau, Nature (London) {\bf 530}, 194 (2016).

\bibitem{Chomaz2016} L. Chomaz, S. Baier, D. Petter, M. J. Mark, F. Wächtler, L. Santos, and F. Ferlaino, Phys. Rev. X {\bf 6}, 041039 (2016).

\bibitem{Schmitt2016} M. Schmitt, M. Wenzel, F. Böttcher, I. Ferrier-Barbut, and T. Pfau, Nature (London) {\bf 539}, 259 (2016).

\bibitem{Wenzel2017} M. Wenzel, F. B\"ottcher, T. Langen, I. Ferrier-Barbut, and T. Pfau,  Phys. Rev. A {\bf 96}, 053630 (2017).

\bibitem{Tanzi2019} L. Tanzi, E. Lucioni, F. Fam\`a, J. Catani, A. Fioretti, C. Gabbanini, R. N. Bisset, L. Santos, and G. Modugno, Phys. Rev. Lett. {\bf 122}, 130405 (2019).

\bibitem{Boettcher2019} F. B\"ottcher, J.-N. Schmidt, M. Wenzel, J. Hertkorn, M. Guo, T. Langen, and T. Pfau, Phys. Rev. X {\bf 9}, 011051 (2019).

\bibitem{Chomaz2019} L. Chomaz, D. Petter, P. Ilzh\"ofer, G. Natale, A. Trautmann, C. Politi, G. Durastante, R. M. W. van Bijnen, A. Patscheider, M. Sohmen, M. J. Mark, and F. Ferlaino, Phys. Rev. X {\bf 9}, 021012 (2019).

\bibitem{Natale2019} G. Natale, R. van Bijnen, A. Patscheider, D. Petter, M. Mark, L. Chomaz, and F. Ferlaino, Phys. Rev. Lett. {\bf 123}, 050402 (2019).

\bibitem{Tanzi2019b} L. Tanzi, S. Roccuzzo, E. Lucioni, F. Fam\`a, A. Fioretti, C. Gabbanini, G. Modugno, A. Recati, and S. Stringari, Nature {\bf 574}, 382 (2019).

\bibitem{Guo2019} M. Guo, F. B\"ottcher, J. Hertkorn, J.-N. Schmidt, M. Wenzel, H. P. B\"uchler, T. Langen, and T. Pfau, Nature {\bf 574}, 386 (2019).


\bibitem{Norcia2021} M. A. Norcia, C. Politi, L. Klaus, E. Poli, M. Sohmen, M. J. Mark, R. N. Bisset, L. Santos, and F. Ferlaino, Nature {\bf 596}, 357 (2021).

\bibitem{Bland2021} T. Bland, E. Poli, C. Politi, L. Klaus, M. A. Norcia, F. Ferlaino, L. Santos, and R. N. Bisset, Phys. Rev. Lett. {\bf 128}, 195302 (2022).

\bibitem{Zhang2019} Y.-C. Zhang, F. Maucher, and T. Pohl,  Phys. Rev. Lett. {\bf 123}, 015301 (2019).

\bibitem{Zhang2021} Y.-C. Zhang, T. Pohl, and F. Maucher, Phys. Rev. A {\bf 104}, 013310 (2021).

\bibitem{Hertkorn2021} J. Hertkorn, J.-N. Schmidt, M. Guo, F. B\"ottcher, K. Ng, S. Graham, P. Uerlings, T. Langen, M. Zwierlein, and T. Pfau, Phys. Rev. Research {\bf 3}, 033125 (2021).

\bibitem{Poli2021} E. Poli, T. Bland, C. Politi, L. Klaus, M. A. Norcia, F. Ferlaino, R. N. Bisset, and L. Santos, Phys. Rev. A {\bf 104}, 063307 (2021).

\bibitem{Gallemi2020} A. Gallem\'i, S. Roccuzzo, S. Stringari, and A. Recati, Phys. Rev. A {\bf 102}, 023322 (2020).

\bibitem{Roccuzzo2020} S. Roccuzzo, A. Gallem\'i, A. Recati, and S. Stringari, Phys. Rev. Lett. {\bf 124}, 045702 (2020).

\bibitem{Cabrera2018} C. R. Cabrera, L. Tanzi, J. Sanz, , B. Naylor, P. Thomas, P. Cheiney, and L. Tarruell, Science {\bf 359}, 301 (2018).

\bibitem{Semeghini2018} G. Semeghini, G. Ferioli, L. Masi, C. Mazzinghi, L. Wolswijk, F. Minardi, M. Modugno, G. Modugno, M. Inguscio, and M. Fattori, Phys. Rev. Lett. {\bf 120}, 235301 (2018).

\bibitem{DErrico2019} C. D'Errico, A. Burchianti, M. Prevedelli, L. Salasnich, F. Ancilotto, M. Modugno, F. Minardi, and C. Fort, Phys. Rev. Research {\bf 1}, 033155 (2019).

\bibitem{Sachdeva2020} R. Sachdeva, M. Nilsson Tengstrand, and S. M. Reimann, Phys. Rev. A {\bf 102}, 043304 (2020).

\bibitem{Sanchez-Baena2020} J. Sánchez-Baena, J. Boronat, and F. Mazzanti, Phys. Rev. A {\bf 102}, 053308 (2020).

\bibitem{Trautmann2018} A. Trautmann, P. Ilzhöfer, G. Durastante, C. Politi, M. Sohmen, M. J. Mark,  and F. Ferlaino, Phys. Rev. Lett. {\bf 121}, 213601 (2018).

\bibitem{Durastante2020} G. Durastante, C. Politi, M. Sohmen, P. Ilzhfer, M. J. Mark, M. A. Norcia, and F. Ferlaino, Phys. Rev. A {\bf 102}, 033330 (2020).

\bibitem{Politi2021} C. Politi, A. Trautmann, P. Ilzh\"ofer, G. Durastante, M. J. Mark, M. Modugno, and F. Ferlaino, Phys. Rev. A {\bf 105}, 023304 (2022).

\bibitem{Bisset2021} R. N. Bisset, L. A. Pe\~na Ardila, and L. Santos, Phys. Rev. Lett. {\bf 126}, 025301 (2021).

\bibitem{Smith2021} J. C. Smith, D. Baillie, and P. B. Blakie, Phys. Rev. Lett. {\bf 126}, 025302 (2021).

\bibitem{Smith2021b} J. C. Smith, P. B. Blakie, and D. Baillie, Phys. Rev. A {\bf 104}, 053316 (2021).

\bibitem{footnote-Modugno} The physical mechanism behind this observation remains unclear. It occurs for large-enough $a_{12}$, under conditions of almost full immiscibility, and hence it does not result from 
the local enhancement of the effective dipolar strength discussed in this paper.

\bibitem{footnote-SM} See the Supplemental Material~(which contains Refs. \cite{Bisset2021,Smith2021,Smith2021b,Eberlein2005,Chomaz2018}), 
where we show further examples of the change of the density profiles of the two components 
when $N_2/N$ is varied, make additional comments concerning the overall superfluidity, 
discuss further details of the  model employed to understand droplet catalyzation, and illustrate the possibility of a supersolid of immiscible droplets.

\bibitem{Legget1970}A. J. Legget, Phys. Rev. Lett. {\bf 25}, 1543 (1970).

\bibitem{Legget1998} A. J. Leggett , J. of Stat. Phys. {\bf 93}, 927 (1998).

\bibitem{Tanzi2021} L. Tanzi, J. G. Maloberti, G. Biagioni, A. Fioretti, C. Gabbanini, and G. Modugno, Science {\bf 371}, 1162 (2021).

\bibitem{Norcia2021b} M. A. Norcia, E. Poli, C. Politi, L. Klaus, T. Bland, M. J. Mark, L. Santos, R. N. Bisset, and F. Ferlaino, arXiv:2111.07768 (2021).

\bibitem{Roccuzzo2022} S. M. Roccuzzo, A. Recati, and S. Stringari, Phys. Rev. A {\bf 105}, 023316 (2022).

\bibitem{footnote-CounterMotion} Counter-motion of the supersolid mixture, which may be triggered by a component-dependent perturbation, is an intriguing topic that will be the 
focus of future studies.

\bibitem{footnote-C} Due to the crossover character of the ID-SS transition, choosing a specific boundary is always somehow arbitrary. In this case, it just serves the purpose of 
highlighting in the ground-state diagram the regions of Figs.~\ref{fig:2}(a) and (b) in which one of the components is expected to be significantly more coherently linked than the other~(see also the discussion of the relative superfluid fraction in~\cite{footnote-SM}). Choosing a slightly different boundary value of ${\cal C}$  does not significantly change the overall phase diagram of Fig.~\ref{fig:3}. A similar phase diagram is recovered 
if we impose the criterion $f_s^{(j)}>0.15$ as the boundary between SS and ID.

\bibitem{Chomaz2018} L. Chomaz, R. M. W. van Bijnen, D. Petter, G. Faraoni, S. Baier, J. H. Becher, M. J. Mark, F. W\"achtler, L. Santos, and F. Ferlaino, 
Nature Phys. {\bf 14}, 442 (2018).

\bibitem{Eberlein2005} C. Eberlein, S. Giovanazzi, and D. H. J. O'Dell, Phys. Rev. A {\bf 71}, 033618 (2005).

\end{thebibliography}

\begin{thebibliography}{99}


\bibitem{Smith2021b} J. C. Smith, P. B. Blakie, and D. Baillie, Phys. Rev. A {\bf 104}, 053316 (2021).

\bibitem{Eberlein2005} C. Eberlein, S. Giovanazzi, and D. H. J. O'Dell, Phys. Rev. A {\bf 71}, 033618 (2005).

\bibitem{Chomaz2018} L. Chomaz, R. M. W. van Bijnen, D. Petter, G. Faraoni, S. Baier, J. H. Becher, M. J. Mark, F. W\"achtler, L. Santos, and F. Ferlaino, 
Nature Phys. {\bf 14}, 442 (2018).

\bibitem{Bisset2021} R. N. Bisset, L. A. Pe\~na Ardila, and L. Santos, Phys. Rev. Lett. {\bf 126}, 025301 (2021).

\bibitem{Smith2021} J. C. Smith, D. Baillie, and P. B. Blakie, Phys. Rev. Lett. {\bf 126}, 025302 (2021).

\end{thebibliography}
\end{document}